\documentclass[aapm,mph,graphicx]{revtex4-1}

\usepackage[mathlines]{lineno}
\modulolinenumbers[5]

\usepackage{hyperref}
\usepackage{xcolor}
\hypersetup{
	colorlinks,
	linkcolor={red!50!black},
	citecolor={blue!50!black},
	urlcolor={blue!80!black}
}
\usepackage{graphicx}
\usepackage[separate-uncertainty = true]{siunitx}  
\usepackage{url}   
\usepackage{amsmath}
\usepackage{amssymb}
\usepackage{booktabs}

\usepackage{xspace} 

\newcommand{\ED}{\ensuremath{n}\xspace}                 
\newcommand{\RED}{\ensuremath{\widehat{n}}\xspace}          
\newcommand{\EDF}{\ensuremath{\nu_i}\xspace}            

\newcommand{\RAC}{\ensuremath{\widehat{\mu}}\xspace}        
\newcommand{\RACS}{\ensuremath{\widehat{\mu}_\textrm{s}}\xspace}     
\newcommand{\RACH}{\ensuremath{\widehat{\mu}_\textrm{h}}\xspace}     
\newcommand{\RACL}{\ensuremath{\widehat{\mu}_\textrm{l}}\xspace}     
\newcommand{\CS}{\ensuremath{\sigma}\xspace}            
\newcommand{\RCS}{\ensuremath{\widehat{\sigma}}\xspace}     
\newcommand{\RCSS}{\ensuremath{\widehat{\sigma}_\textrm{s}}\xspace}  
\newcommand{\RCSH}{\ensuremath{\widehat{\sigma}_\textrm{h}}\xspace}  
\newcommand{\RCSL}{\ensuremath{\widehat{\sigma}_\textrm{l}}\xspace}  
\newcommand{\EAN}{\ensuremath{Z_\textrm{eff}}\xspace}   
\newcommand{\CTNS}{\ensuremath{\xi_\textrm{s}}\xspace}              
\newcommand{\CTNL}{\ensuremath{\xi_\textrm{l}}\xspace}              
\newcommand{\CTNH}{\ensuremath{\xi_\textrm{h}}\xspace}              

\newcommand{\UMETH}{\ensuremath{u_\textrm{meth}(\RED)/\RED}\xspace}      
\newcommand{\UCALIB}{\ensuremath{u_\textrm{calib}(\RED)/\RED}\xspace}      

\begin{document}


\title{Accuracy of image-based electron-density assessment using dual-energy computed tomography} 




\author{Christian M\"ohler$^{1,2}$, Patrick Wohlfahrt$^{3,4}$, Christian Richter$^{3-6}$, Steffen Greilich$^{1,2}$}

\affiliation{$^1$ German Cancer Research Center (DKFZ), Division of Medical Physics in Radiation Oncology, Heidelberg, Germany\\}

\affiliation{$^2$ National Center for Radiation Research in Oncology (NCRO), Heidelberg Institute for Radiation Oncology (HIRO), Heidelberg, Germany\\}

\affiliation{$^3$ OncoRay - National Center for Radiation Research in Oncology, Faculty of Medicine and University Hospital Carl Gustav Carus, Technische Universit\"at Dresden, Helmholtz-Zentrum Dresden - Rossendorf, Dresden, Germany\\}

\affiliation{$^4$ Helmholtz-Zentrum Dresden - Rossendorf, Institute of Radiooncology - OncoRay, Dresden, Germany\\}

\affiliation{$^5$ Department of Radiation Oncology, Faculty of Medicine and University Hospital Carl Gustav Carus, Technische Universit\"at Dresden, Dresden, Germany\\}

\affiliation{$^6$ German Cancer Consortium (DKTK), Dresden, Germany\\}

\author{\footnotesize This is the \textbf{pre-peer reviewed version} of the article entitled ``Methodological accuracy of image-based electron-density assessment using dual-energy computed tomography'', which has been accepted for publication in Medical Physics (\href{http://doi.org/10.1002/mp.12265}{DOI:10.1002/mp.12265}). This article may be used for non-commercial purposes in accordance with Wiley Terms and Conditions for Self-Archiving.}


\date{April 2017}

\begin{abstract}

\textbf{Purpose:} Electron density is the most important tissue property influencing photon and ion dose distributions in radiotherapy patients. Dual-energy computed tomography (DECT) enables the determination of electron density by combining the information on photon attenuation obtained at two different tube voltages. Most algorithms suggested so far use the CT numbers provided after image reconstruction as input parameters, i.e. are imaged-based. To explore the accuracy that can be achieved with these approaches, we quantify the intrinsic methodological and calibration uncertainty of the seemingly simplest approach.

\textbf{Methods:} The core of this approach under study is a one-parametric linear superposition (`alpha blending') of the two DECT images, which is shown to be equivalent to an affine relation between the photon attenuation cross sections of the two x-ray energy spectra. We propose to use the latter relation for empirical calibration of the spectrum-dependent blending parameter. For a conclusive assessment of the electron-density uncertainty, we chose to isolate the purely methodological uncertainty component from CT-related effects such as noise and beam hardening. 

\textbf{Results:} Analyzing calculated spectral-weighted attenuation coefficients, we find universal applicability of the investigated approach to arbitrary mixtures of human tissue with an upper limit of the methodological uncertainty component of 0.15\%, excluding high-$Z$ elements such as iodine.
The proposed calibration procedure is bias-free and straightforward to perform using standard equipment. Testing the calibration on five published data sets, we obtain very small differences in the calibration result in spite of different experimental setups and CT protocols used. Employing a general calibration per scanner type and voltage combination is thus conceivable. 

\textbf{Conclusion:} Given the high suitability for clinical application of the alpha-blending approach in combination with a very small methodological uncertainty, we conclude that further refinement of image-based DECT-algorithms for electron-density assessment is not advisable.
	
\end{abstract}


\maketitle 



%
%

%



\section{Introduction}\label{sec:intro}

A clinical dual-energy computed tomography (DECT) scan comprises two images, corresponding to a map of photon attenuation coefficients of different spectral weighting. While primarily intended for benefits in diagnostic radiological applications \cite{Flohr2006}, DECT has become of particular interest in radiotherapy in recent years \cite{VanElmpt2016}. The combined information in the two images can be used to extract an electron-density and an effective-atomic-number 3D map of the patient \cite{Rutherford1976}. Such quantities can be employed in a physics-based approach to convert CT numbers to the required input quantity of treatment planning systems: electron density for conventional photon therapy; stopping-power ratio (SPR) for proton or heavier ion therapy. In particular for the latter, DECT methods offer the potential to reduce treatment uncertainties \cite{Yang2010,Hunemohr2014a,Hudobivnik2016,Wohlfahrt2016}. For physics-based SPR prediction, an accurate and robust determination of the electron density is crucial, as it is responsible for about 95\% of the variability in SPR. In contrast, determination of an effective atomic number is not necessary for SPR prediction \cite{Farace2014,Mohler2016}.

Various algorithms for the determination of electron density and/or effective atomic number with a clinical DECT scanner have been proposed \cite{Bazalova2008,Saito2012,Landry2013,Bourque2014,Hunemohr2014a,VanAbbema2015}. Most of these algorithms use the CT numbers as provided by the scanner in the two images for the high and low tube voltage as input. This decouples -- to a certain extent -- the source of uncertainty in the CT numbers originating from scanner technology and image reconstruction and the inherent accuracy of the algorithm to derive electron density.

Driven by the demand for high accuracy, the complexity of published algorithms tends to increase, e.g. employing a sophisticated parameterization of the photon attenuation cross section \cite{VanAbbema2015}. This is in contrast to the requirements of clinical implementation, which favors a simple and robust approach in terms of practical calibration, required input parameters or computational demands. A correlation between complexity and practical clinical accuracy is not evident.

We therefore investigated the seemingly simplest of available algorithms, which is a one-parametric linear superposition of the two DECT images, referred to in this manuscript as `alpha blending'. Such an approach was originally empirically postulated by Saito \cite{Saito2012}, therein referred to as `dual-energy subtraction'. Meanwhile, the algorithm is commercially available in a clinical software product (\texttt{syngo.via rho/Z}, Siemens Healthineers, Forchheim, Germany). It was also successfully experimentally validated by H\"unemohr et al.\cite{Hunemohr2014a}, where a mean absolute deviation of measured from reference electron densities of $0.4\%$ was found for tissue substitutes. Studies of other algorithms reported similar results \cite{Bourque2014,VanAbbema2015}, without any clear advantage for one or the other method. 

While such studies comparing reference values and measurements of tissue substitutes are important as a first performance test for a new algorithm, the determined accuracy cannot be easily generalized and translated one-to-one to clinical conditions. On the one hand, the reference electron density of tissue substitutes is itself subject to an uncertainty of often unknown, yet likely not negligible magnitude. On the other hand, results are influenced by CT-related uncertainties which depend on the specific experimental setup and scan protocol (e.g. type and number of used inserts, type/geometry/diameter of phantom, beam hardening correction). The results of such tissue-substitute measurements are thus not directly transferable to a patient scan. For example, the experimental setup -- unlike the patient setup -- is often optimized in some way (e.g. simple geometries, inserts always on central axis).

Consequently, we suggest to separate the sources of uncertainty in the evaluation of algorithms for image-based electron-density determination. Assuming ideal CT numbers as input, we quantified the purely methodological uncertainty of the alpha-blending approach for arbitrary mixtures of human tissue. Secondly, we evaluated the uncertainties of a proposed calibration method.

\section{Methods}

\subsection{Electron-density assessment with dual-energy CT}

The core of the studied approach is a simple parameterization for the electron density relative to water, \RED, linear superposition of the spectrally weighted attenuation coefficient relative to water \RACH (\RACL) for the higher (lower) tube voltage with a single parameter $\alpha$:
\begin{linenomath}
\begin{equation}
	\RED = \alpha \RACH + (1-\alpha) \RACL \, .
	\label{eq:RED_1param}
\end{equation}
\end{linenomath}
We chose this particular representation of equation \ref{eq:RED_1param}, which is in image processing referred to as alpha blending, as it best illustrates the mathematical structure of an affine combination and allows for an elegant form of the derived equations in section \ref{sec:methods_uncertainty}. Furthermore, the correspondence to the calculation of pseudo-monoenergetic CT images is obvious in this form \cite{Kuchenbecker2015}. Please note, that in the case of electron density as studied here, equation \ref{eq:RED_1param} actually corresponds to a weighted subtraction of the two attenuation coefficients, as $\alpha > 1$. Also, equation~\ref{eq:RED_1param} can be transformed into equivalent formulations by transforming the parameter $\alpha$, e.g. $\alpha' = \alpha - 1$ as used by Saito \cite{Saito2012}.

By eliminating \RED via $\RACS = \RED\RCSS$ with $\textrm{s}=\{\textrm{h},\textrm{l}\}$, equation \ref{eq:RED_1param} is equivalent to an affine relation between the relative photon attenuation cross sections, \RCSH and \RCSL:
\begin{linenomath}
\begin{equation}\label{eq:RCS_relation}
	\RCSH = 1/\alpha + (1-1/\alpha) \RCSL \, .
\end{equation}
\end{linenomath}

\subsection{Calibration of $\alpha$}\label{sec:calibration}

The blending parameter $\alpha$ in equation \ref{eq:RED_1param} needs to be calibrated for the system in use (CT scanner, voltage combination, protocol etc.). We propose an empirical calibration with a set of commercially available tissue substitutes, exploiting equation \ref{eq:RCS_relation}, which includes the following steps:
\begin{enumerate}
	\item Scan bone substitutes with known reference electron density \RED (e.g. Gammex 467), using the same CT protocol as for treatment planning, ideally in a phantom of body-like diameter to mimic similar beam hardening conditions. 
	\item Obtain CT numbers \CTNL, \CTNH of inserts in a region of interest.
	\item Calculate $\RCSS = 1000 \cdot (\CTNS-1)/\RED$ for $\textrm{s} = {\textrm{l}, \textrm{h}}$ 
	\item Obtain $\alpha$ via regression to equation \ref{eq:RCS_relation}.
\end{enumerate}

\subsection{Quantification of uncertainty}\label{sec:methods_uncertainty}

The uncertainty in the electron-density determination can be sub-divided into a methodological component and a CT-related component. The latter comprises all deviations of CT numbers from a perfect correlation with photon attenuation coefficients. It consists of a stochastic part (noise) and various sources of systematic error (CT artifacts, CT-number instability in position and time etc.). The methodological component of uncertainty is accordingly defined as the remaining uncertainty, provided ideal CT numbers, i.e. an exact relation of CT numbers to photon attenuation coefficients.

\subsubsection{Methodological uncertainty}

The methodological uncertainty reflects the validity of the parametric form of equation~\ref{eq:RED_1param}. It can conveniently be evaluated by analyzing deviations of calculated spectral-weighted photon attenuation cross sections, \RCSL, \RCSH, from equation \ref{eq:RCS_relation} (which is equivalent to equation \ref{eq:RED_1param}). We calculated such cross sections for chemical elements, as well as reference human tissues \cite{Woodard1986,White1987} and bone substitutes from the \textsc{Gammex Phantom 467} (Sun Nuclear GmbH, Neu-Isenburg, Germany). To do so, energy-dependent elemental cross sections were obtained from the \textsc{NIST XCOM Photon Cross Sections Database} \cite{Berger1998}. For compounds, the elemental cross sections were superimposed, using electron-density fractions calculated from elemental composition (\cite{Woodard1986} and \cite{White1987} for reference human tissues, \cite[Tab. 1]{Hunemohr2014a} for Gammex tissue substitutes). The resulting energy-dependent cross sections of the respective materials were then integrated over two representative x-ray spectra of the \textsc{Somatom Force} dual-source DECT scanner (Siemens Healthineers, Forchheim, Germany). The combinations 70/150Sn kVp (where Sn signifies additional tin filtration) and 100/140 kVp were chosen as limiting cases of spectral separation typically available in clinical DECT scanners. Additional alteration of spectra by an energy-dependent detector efficiency was not considered, as no relevant effect on the resulting methodological uncertainty is expected. \par

The resulting pairs (\RCSH, \RCSL) of relative cross sections were fitted with linear regression to equation \ref{eq:RCS_relation}. With perfect CT numbers, such a regression should actually result in zero residuals if the alpha-blending equation \ref{eq:RED_1param} was perfectly accurate. Inversely, non-zero residuals can be directly ascribed to the methodological uncertainty in the alpha-blending approach. To quantify the latter, residuals in the \RCSH-dimension, $\delta(\RCSH)$, need to be transformed to a corresponding change in electron density, $\delta(\RED)$, which is subsequently interpreted as electron-density error. This is achieved by inserting $\RCSH + \delta(\RCSH)$ instead of $\RCSH$ in equation \ref{eq:RED_1param} (considering in addition $\RAC=\RED\RCS$) and evaluating the difference to the unchanged equation, i.e. with \RCSH inserted, yielding
\begin{linenomath}
\begin{align*}
	\label{eq:REDerror1}
	\delta(\RED)
	&= \{\alpha \RED (\RCSH+\delta(\RCSH)) + (1-\alpha) \RED\RCSL\} - \{\alpha \RED \RCSH + (1-\alpha) \RED \RCSL\} \\
	&= \alpha \RED \delta(\RCSH) \, .
\end{align*}
\end{linenomath}
The relative error in electron density, resulting from an absolute deviation, $\delta(\RCSH)$ from a perfect relation \ref{eq:RCS_relation}, thus results as
\begin{equation}
\label{eq:REDerror1}
\frac{\delta(\RED)}{\RED} = \alpha \delta(\RCSH) \, .
\end{equation}
The derivation of a general methodological standard uncertainty \cite{JCGM2008} from a distribution of such calculated errors requires prior knowledge or assumptions on the relative abundance of the substances in the specific imaging situation. However, for a certain class of substances (e.g. human tissue only), an upper limit of the methodological standard uncertainty is always given by the maximum absolute error of the substances within the class, i.e.
\begin{linenomath}
\begin{equation}
	\label{eq:umeth}
	\frac{u_\textrm{meth}(\RED)}{\RED} < \max_{i}{\left| \frac{\delta_i(\RED)}{\RED} \right|}
\end{equation}
\end{linenomath}

\subsubsection{Calibration uncertainty}

The calibration uncertainty was assessed by re-analyzing data from a previous study in our group \cite{Hunemohr2014a} and three more publications \cite{Tsukihara2013,Landry2011,Landry2013}, providing measured CT numbers for the 80/140Sn kVp voltage combination of the CT scanner \textsc{Somatom Definition Flash} (Siemens Healthineers). Starting with the provided CT numbers in step (iii) of the calibration procedure (section \ref{sec:calibration}), corresponding values for $\alpha$ were obtained independently. The coefficient of variation (i.e. standard deviation divided by the mean) was used as an estimate for the relative standard uncertainty in $\alpha$, $u(\alpha)/\alpha$. The relative standard uncertainty associated to the calibration procedure, $u_\textrm{calib}(\RED)/\RED$, follows via common propagation of uncertainty in equation \ref{eq:RED_1param} as
\begin{linenomath}
\begin{equation}
	\label{eq:REDerror2}
	\frac{u_\textrm{calib}(\RED)}{\RED} = \frac{1}{\RED}\left| \frac{\partial \RED}{\partial \alpha} \right| u(\alpha) = |1-\RCSL| \frac{u(\alpha)}{\alpha}\,.  
\end{equation}
\end{linenomath}

\section{Results}

\subsection{Methodological uncertainty}\label{sec:applicability}

\subsubsection{Reference human tissues}

\begin{figure}
	\centering
	\includegraphics[width=\textwidth]{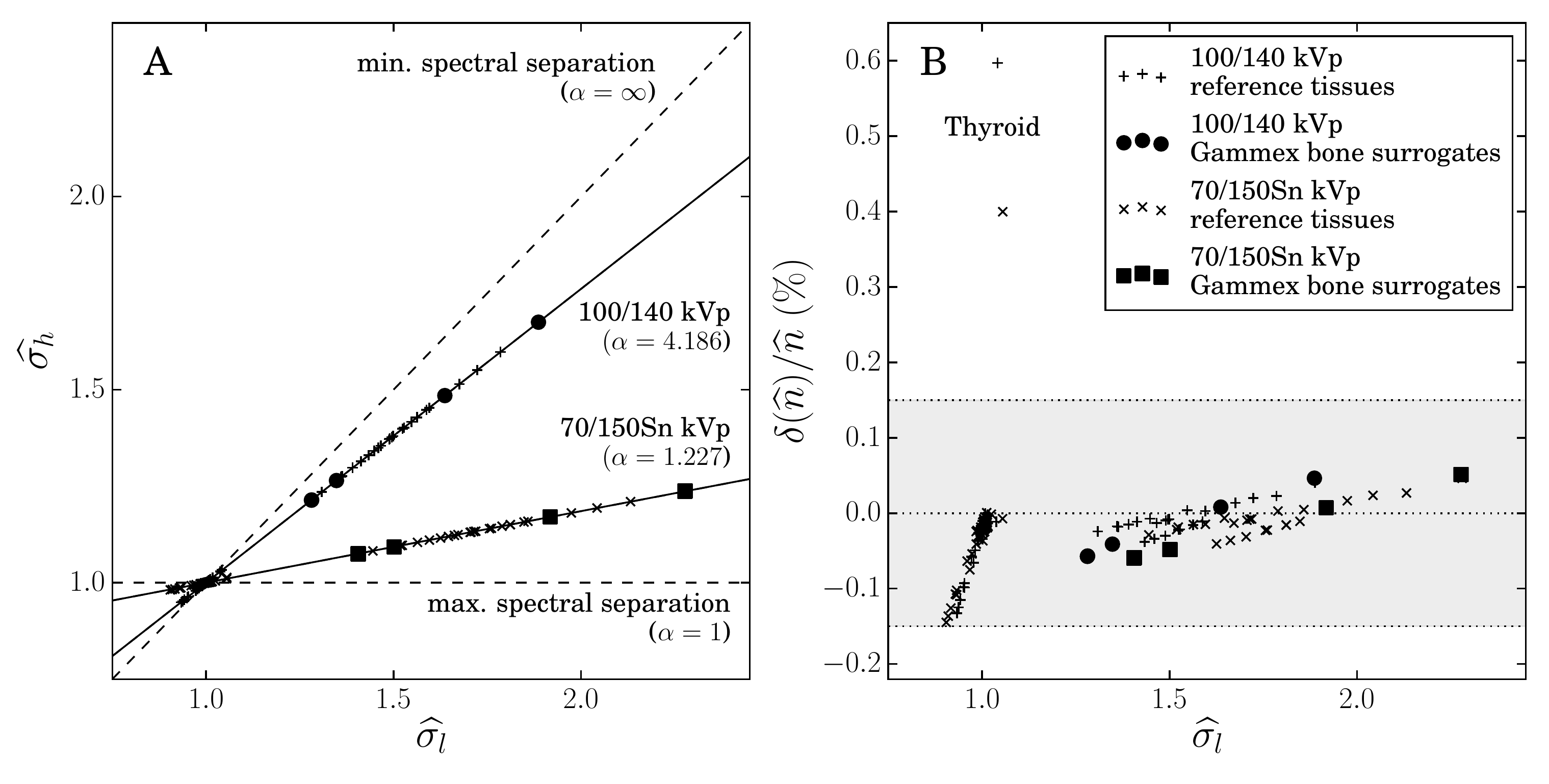}
	\caption{A: Affine functional relationship between calculated spectral-weighted relative cross sections of reference tissues (+, x) and the Gammex bone substitutes Inner bone, B200, CB30, CB50 and Cortical bone ($\circ$, $\square$) for the two investigated spectrum combinations, respectively. The data points for Inner bone and B200 (lowest \RCSL) are overlapping due to very similar elemental composition. Separate linear regression of the reference tissues and bone substitutes to equation \ref{eq:RCS_relation} result in negligible differences in $\alpha$ ($\delta(\alpha) <0.001$), making the respective fit lines indistinguishable. B: Relative error in the assessment of electron density as obtained from the residuals, $\delta(\RCSH)$, to the fit curves in panel A and the respective fitted $\alpha$ via equation \ref{eq:REDerror1}.}
	\label{fig:1}
\end{figure}

The calculated cross sections are described with very good agreement by a linear function as required for the validity of equation \ref{eq:RED_1param} (figure \ref{fig:1}). Only a small trend of deviation is observed in the soft-tissue and bone-tissue region separately, yielding a maximum error of $\delta(\RED)/\RED = 0.15\%$ for low \RCS (fatty tissues), excluding thyroid (discussed in section \ref{sec:highZ}). The distribution of errors is very similar for different spectral separation. With increasing spectral separation, larger residuals $\delta(\RCSH)$ in the linear regression are compensated by a decreasing $\alpha$ according to equation \ref{eq:REDerror1}. The behaviour of voltage combinations of intermediate spectral separation was found to be within the limits of the two cases displayed in figure \ref{fig:1} (data not shown). Consequently, an upper limit of the methodological standard uncertainty, $\UMETH < 0.15\%$, can be stated according to equation \ref{eq:umeth}, independently of the used spectrum combination.

\subsubsection{Influence of high-Z elements}\label{sec:highZ}

\begin{figure}
	\centering
	\includegraphics[width=\textwidth]{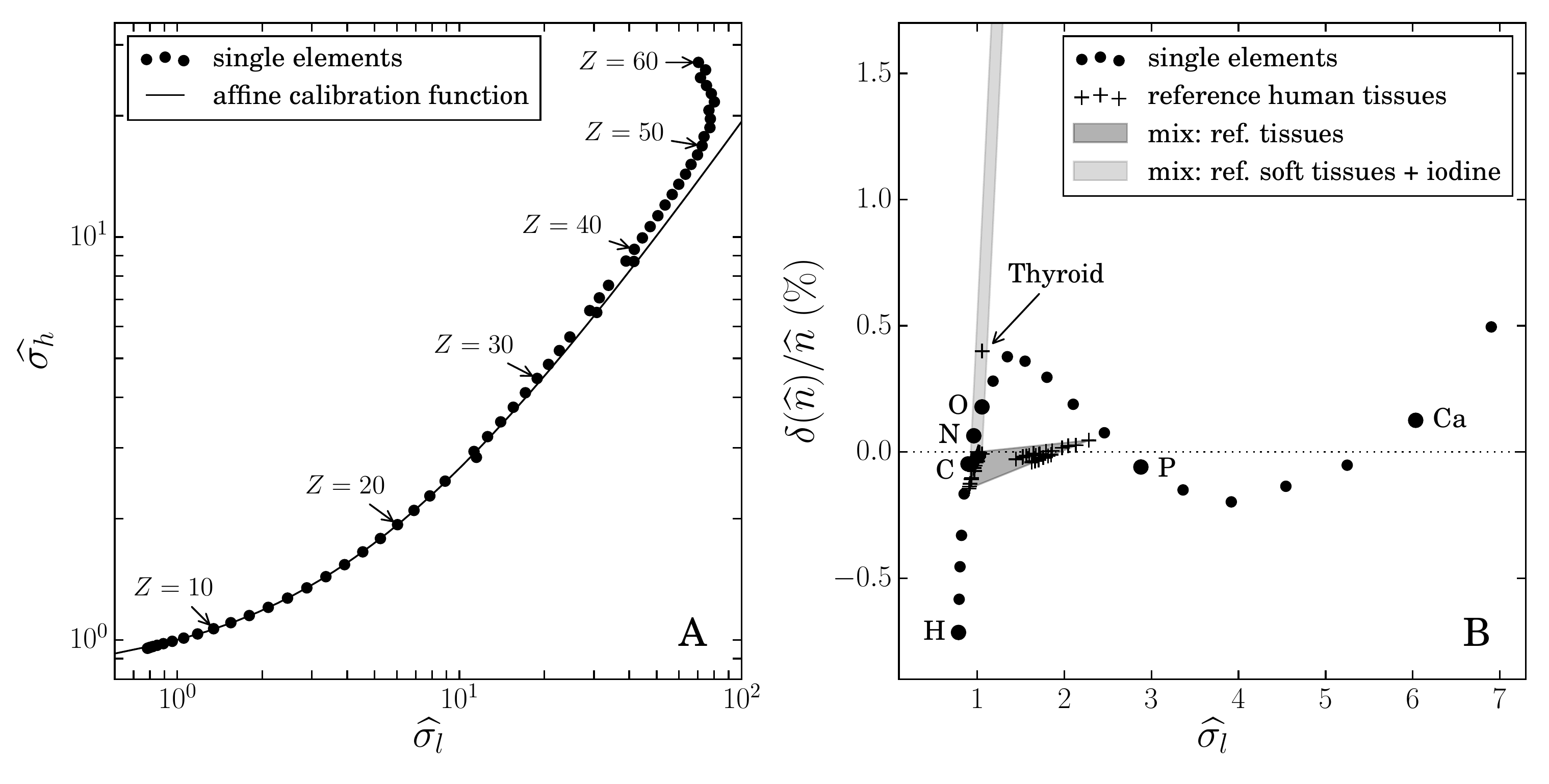}
	\caption{A: Calculated cross sections for the 70/150Sn kVp voltage combination for chemical elements up to $Z=60$. K edges of the photoelectric effect cause a strong deviation from the affine calibration function for high $Z$. B: Corresponding error in the electron-density assessment for single elements (displayed up to $Z=21$) and reference human tissues. All possible mixtures of selected base materials are contained in their convex hull, which is represented as a shaded area (e.g. grey area for arbitrary mixtures of reference tissues excl. thyroid). Thyroid is contained in the convex hull of reference soft tissues and iodine.}
	\label{fig:2}
\end{figure}

Thyroid is the only reference tissue containing a relevant amount ($\approx 0.1\%$ by mass) of a high-$Z$ element (iodine, $Z=53$). The positive error of about $0.5\%$ for thyroid (figure~\ref{fig:1}) can be explained by the influence of the photoelectric effect; the position of the K edge shifts to higher energy for increasing $Z$ (e.g. K edge for iodine at $E=33$ keV) and thus starts to affect the attenuation of CT spectra ($\approx 20-150$ keV). For sufficiently large $Z$, the absorption of the lower-voltage spectrum gets significantly affected by the sudden drop at the K edge, leading to a slower increase, and eventually a decrease of \RCSL with increasing $Z$. At the same time, the absorption of the higher-voltage spectrum (filtered for lower energies) is not or at least less affected. Consequently, an increasingly strong positive divergence from the affine relation between relative cross sections, $\delta(\RCSH) > 0$, is observed for high $Z$ (figure \ref{fig:2}, A), which leads to a positive bias in the determined electron density, $\delta(\RED) > 0$, according to equation \ref{eq:REDerror1}.

To better understand the specific electron-density error for a chemical compound or mixture containing a high-$Z$ element, it is important to analyze mixing properties of relative cross sections. Due to the linear structure of the superposition of relative cross sections, all possible mixtures of a set of base materials in the (\RCSL, \RCSH)-plane form their convex hull. This is actually in strict analogy to the mixing behavior of the stopping number and cross section, as investigated in detail by M\"ohler et al.\cite{Mohler2016}. A mathematical derivation can be found there, but is not relevant for the understanding of the problem discussed here. As illustrated in figure \ref{fig:2}B, thyroid is contained in the convex hull of soft tissue and iodine. In this way, already a small trace of a high-$Z$ element causes a substantial positive bias, of in this case about $0.5\%$, in electron density.

\subsection{Calibration uncertainty}\label{sec:results_calibration}

The blending parameters obtained from published data sets show very little variation, in spite of a variety of experimental configurations and CT protocols used (table \ref{tab:calibration}). With a mean and standard deviation of $\alpha = 1.457 \pm 0.017$, the relative uncertainty in $\alpha$ is estimated as $u(\alpha)/\alpha = 1.2\%$. Defining $\Delta_{\RCSL}^\textrm{max} = \max_i{|1-\RCSL{}_{,i}|}$ for a given tissue group, an upper limit of the calibration uncertainty can be calculated using equation \ref{eq:REDerror2}. For soft (bony) tissue with $\Delta_{\RCSL}^\textrm{max} = 0.05 (0.7)$, we get $\UCALIB < 0.06\% (0.8\%)$.

Of the parameters listed in table \ref{tab:calibration}, only the phantom diameter (head vs. abdomen) showed a significant impact on the variability in $\alpha$ ($p=0.007$), with the head setup leading to smaller $\alpha$-values than the abdomen setup. This is most likely explained by beam hardening effects, as no beam hardening correction for bones was applied in the cited studies.

\begin{table}
	\caption{\small Datasets used in the quantification of the calibration uncertainty. CT data was obtained with the 80/140Sn kVp voltage combination of a Siemens \textsc{Somatom Definition Flash} CT scanner. Different container phantoms for the inserts made of solid water (SW) or acrylic glass (PMMA) with diameters typical for head (H) or abdomen (A) were used. The listed parameters were tested individually for statistical significance of their influence on the variation in $\alpha$ by analysis of variance (ANOVA) ($p$-values listed below the parameter name). Only the bone-tissue substitutes of the respective phantom were used for the regression, while additional inclusion of soft-tissue substitutes did not alter the resulting $\alpha$ in the four significant digits shown. }
	\begin{tabular}{@{}lllllllll}
		\hline\hline
		ID & author & \multicolumn{2}{l}{inserts} & \multicolumn{2}{l}{container phantom} & \multicolumn{2}{l}{reconstruction} & $\alpha$ \\
		&           & vendor & model & material & diameter & kernel & slice & \\
		& $p=0.37$ & $p=0.50$ & $p=0.75$ & $p=0.44$ & $p=0.007$* & $p=0.32$ & $p=0.77$ &  \\ 
		\hline
		C1 & \cite{Tsukihara2013} & CIRS & 062 & SW & $\varnothing \approx \SI{30}{cm}$ (A) & D30 & \SI{5}{mm} & 1.465 \\
		C2 & \cite{Tsukihara2013} & CIRS & 062 & SW & $\varnothing \,\SI{18}{cm}$ (H) & D30 & \SI{5}{mm} & 1.433 \\
		C3 & \cite{Hunemohr2014a} & Gammex & 467 & PMMA & $\varnothing \,\SI{16}{cm}$ (H) & D30 & \SI{2}{mm} & 1.442 \\
		C4 & \cite{Landry2011} & Gammex & 465 & SW & $\varnothing \,\SI{33}{cm}$ (A) & B10 & \SI{3}{mm} & 1.472 \\
		C5 & \cite{Landry2013} & Gammex & 467 & SW & $\varnothing \,\SI{33}{cm}$ (A) & n/a & \SI{3}{mm} & 1.475 \\
		\hline\hline
	\end{tabular}
	\label{tab:calibration}
\end{table}
\normalsize

\section{Discussion}

\subsection{Methodological uncertainty}

The methodological uncertainty, $\UMETH < 0.15\%$, for tissue excl. thyroid and the error, $\delta(\RED)/\RED \approx 0.5\%$ for thyroid are very similar to the ones of the `stoichiometric method' in \cite{Bourque2014}. Their theoretical analysis resulted in a maximum absolute deviation of determined from reference electron densities of $0.2\%$ for reference human tissues excl. thyroid and a positive deviation of $0.6\%$ for thyroid (figure 6 in Bourque et al.\cite{Bourque2014}). Furthermore, the distribution of residuals in figure \ref{fig:1}B is remarkably similar to the residuals observed in a fitting step of the ratio of attenuation coefficients to \EAN in figure 7 of Landry et al.\cite{Landry2013}. These similarities are made plausible by the fact that these approaches use a single energy-independent effective atomic number, which requires a cross section parameterization of the type of equation \ref{eq:cross_section_basic} (appendix \ref{sec:framework}). From this parameterization, equation \ref{eq:RED_1param} can be rigorously derived (appendix \ref{sec:proof2}). Every approach, in which an effective atomic number is defined (table \ref{tab:algorithms}, M3-6) thus implicitly recognizes the validity of equation \ref{eq:RED_1param}. It is natural that such an algorithm would include a fitting step at some point with residuals to reference tissues that are (in the best case) of similar structure and magnitude as shown in figure \ref{fig:1}B. This also means that the use of a parameterization not respecting the form of equation \ref{eq:cross_section_basic} to determine a unique \EAN (as in M3-6) is inconsistent.

\begin{table}
	\caption{\small Selected published algorithms for image-based determination of relative electron density, \RED, and/or effective atomic number, \EAN. M1 and M2 use the alpha-blending equation \ref{eq:RED_1param} and thus do not require a parameterization for the cross section (parameterization in M2 only for \EAN). M3-6 use parameterizations, which do not respect the condition of equation \ref{eq:cross_section_basic}. Some methods require the input of x-ray spectra (`spec'), while others are based on calibration (`cal').}
	\footnotesize
	\begin{tabular}{@{}lllll}
		\hline\hline
		ID & reference                    & \RED/\EAN               & kind & parameterization  \\
		\hline
		M1 & \cite{Saito2012} & \RED & cal & - \\
		M2 & \cite{Hunemohr2014a}  & \RED $\rightarrow$ \EAN & cal & $a(E) + b(E) Z^m$ (only for \EAN) \\
		M3 & \cite{Bazalova2008}    & \EAN $\rightarrow$ \RED & spec   & $Z^4 F(E,Z) + G(E,Z)$                 \\
		M4 & \cite{Landry2013}      & \EAN                    & cal & \cite{Rutherford1976} \\
		M5 & \cite{Bourque2014}     & \EAN $\rightarrow$ \RED & cal & $\sum_k{a_k Z^k}$ \\
		M6 & \cite{VanAbbema2015}      & \EAN $\rightarrow$ \RED & spec & \cite{Jackson1981} \\
		\hline\hline
	\end{tabular}\\
	\label{tab:algorithms}
\end{table}
\normalsize

\subsection{Calibration uncertainty}

The determined calibration uncertainty can be considered a rather conservative estimate, as it was derived from a spread of data using differently optimized setups and CT protocols. Still, the calibration uncertainty can be safely neglected for soft tissue ($\UCALIB < 0.06\%$), due to the similarity to water in elemental composition ($\RCS \approx 1$). For bones, the calibration uncertainty ($\UCALIB < 0.8\%$) might be further reduced by a body-site-specific calibration or the use of a CT reconstruction kernel with beam hardening correction for bone. On the other hand, the calibration result appears to be robust against changes in setup or CT protocol with only a small influence of the phantom/patient diameter. A general calibration per scanner type and voltage combination is thus conceivable, simplifying clinical implementation.

The calibration procedure (section \ref{sec:calibration}) requires one CT scan of only a few bone tissue substitutes and one linear regression directly to equation \ref{eq:RCS_relation}. It can thus be considered fairly more straightforward and presumably more robust than the calibration procedure proposed by Saito \cite{Saito2012}, which involves optimizing a correlation coefficient in a multi-step regression. Furthermore, the calibration result is independent of whether tissue substitutes or reference human tissues are used (figure \ref{fig:1}). An intermediate stoichometric calibration step like the one in the ``RTM method'' of Landry et al. \cite{Landry2013} is thus not required. The calibration should be performed using only bone-tissue substitutes ($\RCS \gg 1$), as soft-tissue substitutes do not significantly alter the calibration outcome in the best case. In the worst case (i.e. without a sufficient number of higher-\RCS calibration materials) they might even destabilize the fit. For the same reason, a refinement of the calibration in the soft-tissue region e.g. with a slightly different slope (or $\alpha$) -- as might be suggested by the observed trend in figure \ref{fig:1}B -- is not practical. 

Quantification of an uncertainty budget that includes deviations of CT numbers from actual photon attenuation coefficients for all clinical situations conceivable, would, if feasible and appropriate at all, require extensive experimental investigation. This was clearly beyond the scope and intention of this work.

\section{Conclusion}

Accurate and robust electron-density imaging with a clinical DECT scanner is achieved using one-parametric alpha blending of the two CT images. The algorithm is generally applicable to arbitrary mixtures of human tissue with a maximum methodological uncertainty below $0.15\%$, excluding high-$Z$ elements. Calculation of an effective atomic number is possible as well, but not necessary. Empirical calibration of the blending parameter is straightforward, bias-free and robust against the specific setup and CT protocol in use. The alpha-blending approach can thus be considered highly suitable for clinical application. Further improvement of the basic algorithm is neither possible without certain loss of universal applicability concerning mixtures (appendix \ref{sec:framework}); nor reasonable comparing the very small methodological uncertainty with other sources of uncertainty in the delivery of radiotherapy. Beside factors that improve CT-number accuracy and consequently image-based assessment of electron density, our study does also not address the potential benefit of sinogram-based methods.

\begin{acknowledgments}

This work was partially funded by the National Center for Radiation Research in Oncology (NCRO) and the Heidelberg Institute for Radiation Oncology (HIRO) within the project `translation of dual-energy CT into application in particle therapy'. CR is funded by the German Federal Ministry for Education and Research (Grant No. BMBF-03Z1N51).

\end{acknowledgments}

\appendix

\section{Notation and concepts}\label{sec:methods}

\subsection{Photon attenuation and CT numbers}\label{sec:notation}

The photon attenuation coefficient, $\mu$, factorizes into electron density, \ED, and electronic cross section, \CS, such that
\begin{linenomath}
\begin{equation}\label{eq:RAC_RED_RCS}
	\mu = \ED\CS \, .
\end{equation}
\end{linenomath}
The CT number, $\xi$, in Hounsfield units (HU), as provided by a CT scanner, is linked to the effective, i.e. spectral-weighted, attenuation coefficient relative to water, \RACS, via
\begin{linenomath}
\begin{equation}\label{eq:CTnumber}
	\xi = (\RACS - 1) \cdot \SI{1000}{HU} \, .
\end{equation}
\end{linenomath}
Here, a hat on a variable's symbol marks a quantity $x$ normalized by the same quantity for water $x^\textrm{w}$, i.e. $\widehat{x} = x/x^\textrm{w}$. The index $\textrm{s}$ signifies spectral weighting in the form $x_\textrm{s} = \int{S(E) x(E) \, \textrm{d}E}$,
where the detected spectrum $S(E)$, normalized such that $\int{S(E) \, \textrm{d}E} = 1$, contains the emitted x-ray spectrum and energy-dependent detector properties such as efficiency.

\subsection{Cross section parameterization}

Many specific parameterizations of the photon absorption cross section, $\CS(E, Z)$, in terms of photon energy, $E$, and atomic number of the absorber, $Z$, have been proposed for the energy regime of medical imaging, i.e. $E<1$ MeV \cite{Rutherford1976,Alvarez1976,Jackson1981}. Among these, an important general shape is given by
\begin{linenomath}
\begin{equation}\label{eq:cross_section_basic}
	\sigma(E,Z) = A(E) + B(E)C(Z) \, ,
\end{equation}
\end{linenomath}
corresponding to eq. 4.6 in Jackson and Hawkes\cite{Jackson1981}. The important feature of this parameterization is the separation of $E$- from $Z$-dependence in the second term. A power law,
\begin{linenomath}
\begin{equation}\label{eq:Z_to_the_m}
	C(Z) = Z^m \, ,
\end{equation}
\end{linenomath}
is often used in literature, where $m \in [3,4]$ if only the photoelectric effect is considered.

\subsection{Effective atomic number}

From the attenuation sum rule for chemical compounds or mixtures, $\mu = \sum_i \mu_i$, and equation \ref{eq:RAC_RED_RCS}, the energy-dependent cross section of a composite material with electron-density fractions, $\EDF = \ED_i/\ED$, results as
\begin{linenomath}
\begin{equation}\label{eq:CS_compound}
	\sigma(E) = \sum\nolimits_i \nu_i \sigma(E, Z_i) \equiv \sigma(E, \EAN) \, .
\end{equation}
\end{linenomath}
On the right-hand side, the effective atomic number, \EAN, is defined as the virtual decimal atomic number of a hypothetical one-atomic element that would express the same cross section at a given photon energy as the compound in consideration, based on a given cross section parameterization, $\sigma(E,Z)$. It thus reduces the number of variables to describe photon interaction with a mixture. A cross section parameterization in the form of equation \ref{eq:cross_section_basic} is a necessary condition for the definition of a single energy-independent effective atomic number \cite{Jackson1981}. The definition \ref{eq:CS_compound} then translates to
\begin{linenomath}
\begin{equation}\label{eq:definition_EAN_specific}
	C(\EAN) = \sum\nolimits_i \EDF C(Z_i) \, .
\end{equation}
\end{linenomath}
In order to solve for \EAN, $C(Z)$ has to be invertible. Using equation \ref{eq:Z_to_the_m} leads to the well-known and frequently used equation
\begin{linenomath}
\begin{equation}\label{eq:common_definition_EAN}
	\EAN = \left( \sum\nolimits_i \EDF Z_i^m \right)^{1/m} \, .
\end{equation}
\end{linenomath}

\section{Mathematical framework}\label{sec:framework}

\begin{figure}
	\centering
	\includegraphics[width=0.6\textwidth]{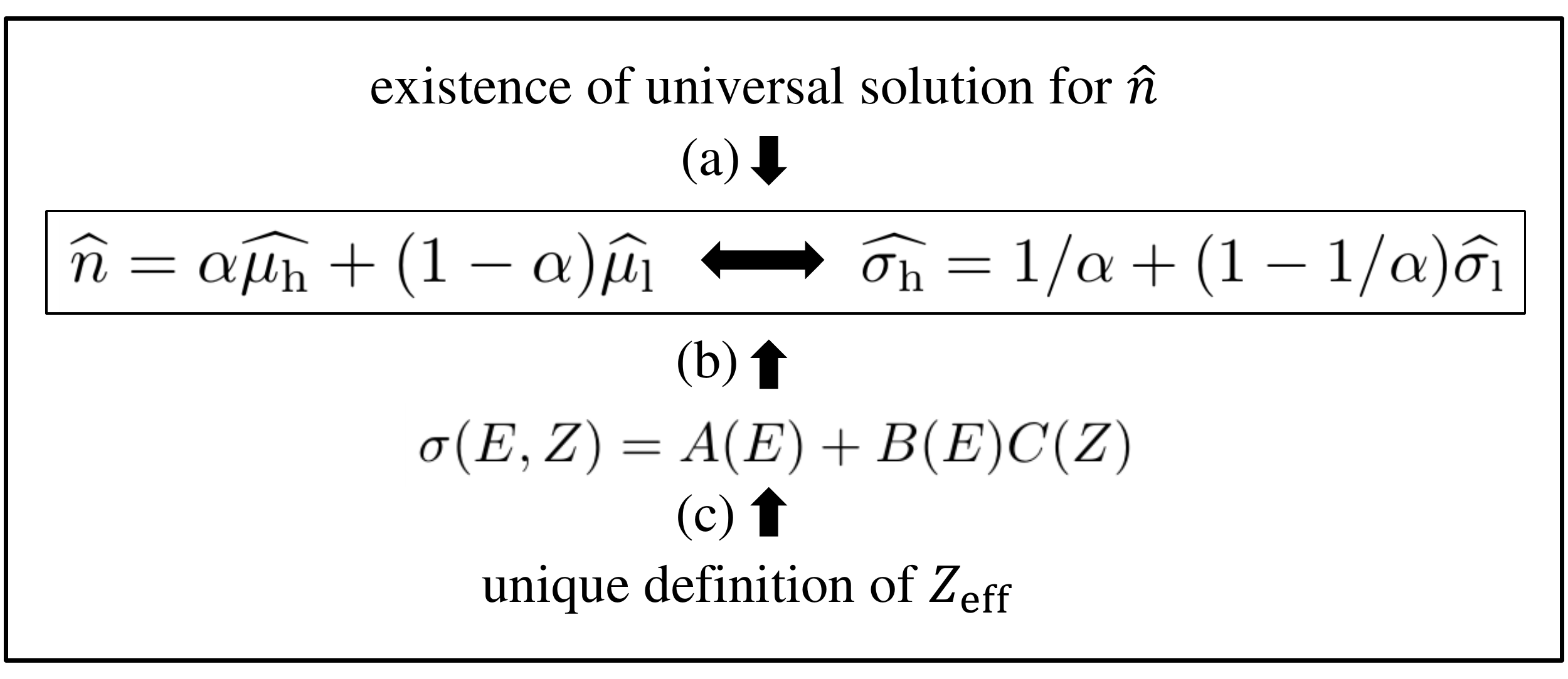}
	\caption{Visualization of mathematical implications.}
	\label{fig:flow_chart}
\end{figure}

The alpha-blending equation \ref{eq:RED_1param} for electron-density assessment can be derived rigorously by requiring a universal solution for the electron density, meaning that the solution is valid for any single chemical element as well as any arbitrary compound (figure \ref{fig:flow_chart}(a), proof in appendix \ref{sec:proof1}). Alternatively, the central equations can be derived from the given general functional dependency of the cross section on $Z$ and $E$ (figure \ref{fig:flow_chart}(b), proof in appendix \ref{sec:proof2}). A cross section parameterization of the given form is a necessary requirement for the definition of a unique, energy-independent effective atomic number (figure \ref{fig:flow_chart}(c), Jackson and Hawkes\cite{Jackson1981}).

\subsection{Derivation of the electron-density equation \ref{eq:RED_1param} from the requirement of a universal solution}\label{sec:proof1}

DECT provides two CT numbers and thus via equation \ref{eq:CTnumber} two relative attenuation coefficients, \RACL and \RACH. Writing equation \ref{eq:RAC_RED_RCS} for these, we get a system of two equations,
\begin{linenomath}
\begin{equation}\label{eq:basic_system1}
	\RACS = \RED\RCSS \, ,
\end{equation}
\end{linenomath}
with $\textrm{s}=\{\textrm{h},\textrm{l}\}$. We require now a unique and universal solution for the electron density, in the sense that it provides the correct result for single chemical elements, as well as all kinds of compounds and mixtures without any pre-knowledge on the substance. As equation \ref{eq:basic_system1} contains three unknowns, \RED, \RCSH and \RCSL, we need to find a unique function
\begin{linenomath}
\begin{equation}
	f: \RCSL \longrightarrow \RCSH \, ,
\end{equation}
\end{linenomath}
connecting the relative cross sections. The function $f$ has to apply for single elements with elemental cross sections $\RCSS{}_{,i}$,
\begin{linenomath}
\begin{equation}
	f(\RCSL{}_{,i}) = \RCSH{}_{,i} \quad \forall i \, ,
	\label{eq:condition_elements}
\end{equation}
\end{linenomath}
as well as for all possible mixtures, for which we know from the attenuation sum rule that $\RCSS = \sum_i \EDF\RCSS{}_{,i}$, leading to the condition
\begin{linenomath}
\begin{equation}
	f\left( \sum_i \EDF\RCSL{}_{,i} \right) = \sum_i \EDF\RCSH{}_{,i} \, .
	\label{eq:condition_mixtures}
\end{equation}
\end{linenomath}
Substituting equation \ref{eq:condition_elements} into equation \ref{eq:condition_mixtures} yields
\begin{linenomath}
\begin{equation}
	f\left( \sum_i \EDF\RCSL{}_{,i} \right) = \sum_i \EDF f\left( \RCSL{}_{,i} \right)
	\label{eq:condition_affine}
\end{equation}
\end{linenomath}
With $\sum_i \EDF\RCSL{}_{,i}$ being an affine combination ($\sum_i \EDF = 1$), equation \ref{eq:condition_affine} states that $f$ is an affine function. Now, a universal solution for the electron density should be true for water in particular, yielding the additional constraint $f(\RCSL=1) = 1$. Based on this, a specific parameterization of the affine function $f$ is given by equation \ref{eq:RCS_relation} with one parameter $\alpha \in \mathrm{R}\backslash 0$. Using this function in the original system of equations, equation \ref{eq:basic_system1}, yields the solution of equation \ref{eq:RED_1param}. The affinity of the cross section and thus the central equation \ref{eq:RED_1param} is thus a necessary condition for the existence of a universal solution for the electron density.

\subsection{Derivation of the electron-density equation \ref{eq:RED_1param} from an assumption on the cross section parameterization \ref{eq:RCS_relation}}\label{sec:proof2}

With the definitions in appendix \ref{sec:notation} and equation \ref{eq:CS_compound}, the relative cross section for a compound writes
\begin{linenomath}
\begin{equation}\label{eq:RCS_compound}
	\RCSS = \frac{1}{\sigma_\textrm{s}^\textrm{w}} \int{\textrm{d}E \, S(E) \sum\nolimits_i{\EDF \sigma(E, Z_i)}}
\end{equation}
\end{linenomath}
Using the cross section parameterization of type \ref{eq:cross_section_basic}, we get
\begin{linenomath}
\begin{equation}\label{eq:basic_system2}
	\RCSS = \widetilde{a_\textrm{s}} + \widetilde{b_\textrm{s}} \sum\nolimits_i{\EDF C(Z_i)} \, ,
\end{equation}
\end{linenomath}
where $\widetilde{a_\textrm{s}} \equiv a_\textrm{s}/\sigma_\textrm{s}^\textrm{w}$ and $\widetilde{b_\textrm{s}} \equiv b_\textrm{s}/\sigma_\textrm{s}^\textrm{w}$ contain the dependence on the energy spectrum. With equations \ref{eq:basic_system1} and \ref{eq:basic_system2}, we have now a fully determined system of equations with the two unknowns \RED and $\sum_i{\EDF C(Z_i)}$. The latter equals $C(\EAN)$ according to equation \ref{eq:definition_EAN_specific}, but there is no need to specify the function $C$ at this point. The system of equations is now readily solved for the relative electron density, obtaining
\begin{linenomath}
\begin{equation}
	\RED = \alpha \RACH + \beta \RACL \, .
	\label{eq:RED_2param}
\end{equation}
\end{linenomath}
The newly introduced parameters $\alpha$, $\beta$ are functions of the $\widetilde{a_\textrm{s}}$ and $\widetilde{b_\textrm{s}}$, the shape of which is easily derived, but of no further interest here. The constraint for the water calibration point $1 = \alpha + \beta$ again eliminates one of the parameters and leads to the final form of equation \ref{eq:RED_1param}.

NB: The system of equations \ref{eq:basic_system1} and \ref{eq:basic_system2} could also be solved for \EAN, as shown in \cite[section 2.2]{Hunemohr2014a}. For this, the $Z$-dependence of the cross section needs to be specified, e.g. $C(Z) = Z^{3.1}$. A parameterization of the $E$-dependence in equation \ref{eq:cross_section_basic} is not necessary, as it is absorbed in the calibration parameters.



\end{document}